\documentstyle[12pt]{article}
\newcommand{\half}{\frac{1}{2}}
\newcommand{\G}{{\open C}[W(R)]}
\newcommand{\R}{{R}}
\newcommand{\be}{\begin{equation}}
\newcommand{\ee}{\end{equation}}
\newcommand{\bee}{\begin{eqnarray}}
\newcommand{\eee}{\end{eqnarray}}
\newcommand\nn{\nonumber \\}
\newcommand{\di}{\frac {\partial} {\partial x_i}}
\newcommand\defeq{\stackrel {def}{=}}
\newcommand{\x}{{\vec x}}
\newcommand{\yy}{{\vec y}}
\newcommand{\cc}{{\vec c}}
\newcommand{\vv}{{\vec v}}

\font\frtnfr=eufm10   scaled\magstep1
\font\twlfr=eufm10
\font\tenfr=eufm10

\newfam\frfam
\textfont\frfam=\frtnfr
\scriptfont\frfam=\twlfr
\scriptscriptfont\frfam=\tenfr
\def\fr{\fam\frfam}

\font\frtnopen=msbm10  scaled\magstep2
\font\twlopen=msbm10
\font\tenopen=msbm10

\newfam\openfam
\textfont\openfam=\frtnopen
\scriptfont\openfam=\twlopen
\scriptscriptfont\openfam=\tenopen
\def\open{\fam\openfam}

\font\frtnsf = cmss12 scaled\magstep1
\font\twlsf = cmss10
\font\tensf = cmss9

\newfam\Scfam
\textfont\Scfam = \frtnsf
\scriptfont\Scfam = \twlsf
\scriptscriptfont\Scfam = \tensf

\begin{document}
\renewcommand{\theequation}{\arabic{equation}}
\bibliographystyle{nphys}


\sloppy
\title
 {
      \hfill{\normalsize\sf FIAN/TD/98-25}    \\
            \vspace{1cm}
      Supertraces on the Superalgebra of Observables of
      Rational Calogero Model based on the Root System
 }
\author
 {
   S.E.Konstein
          \thanks
             {E-mail: konstein@td.lpi.ac.ru}
          \thanks
             {This work is supported by the Russian Fund for Basic
               Research,
              Grants 96-02-17314 and 96-15-96463.}
  \\
               {\small \phantom{uuu}}
  \\
           {\it {\small} I.E.Tamm Department of Theoretical Physics,}
  \\
               {\it {\small} P.N.Lebedev Physical Institute,}
  \\
         {\it {\small} 117924, Leninsky Prospect 53, Moscow, Russia.}
 }
\date{ }
\maketitle
\begin{abstract}
It is shown that the superalgebra $H_{W(\R)}$ of observables of the
rational Calogero model based on the root system $\R$ possesses
$Q_\R$ supertraces, where $Q_\R$ is the number of conjugacy classes
of the Coxeter group $W(\R)$ generated by the root system $\R$
which have no eigenvalue $-1$.
\end{abstract}

\newpage

\section{The superalgebra of observables.}
The superalgebra $H_{W(\R)}$ of observables
of the rational Calogero model based on the root system $\R$
is defined in the following way.

For any nonzero $\vv \in V={\open R}^N$ define the reflections
$R_\vv$ as follows:
\be\label{ref}
R_\vv (\x)=\x -2 \frac {(\x,\,\vv)} {(\vv,\,\vv)} \vv \qquad
\mbox{ for any }\x \in V.
\ee
Here $(\cdot,\cdot)$ stands for the inner product in
$V$: $(\x,\,\yy)=\sum_{i=1}^N x_i y_i$, where the $x_i$ are
the coordinates of vector $\x$: $x_i\defeq (\x,\,\vec e_i)$,
and the vectors $\vec e_i$ constitute an orthonormal basis in $V$:
$(\vec e_i,\, \vec e_j)=\delta_{ij}$.
The reflections (\ref{ref}) have the following properties
\be\label{prop}
R_\vv (\vv)=-\vv,\qquad R_\vv^2 =1,\qquad
({R}_\vv (\x),\,\vec u)=(\x,\,{R}_\vv (\vec u)),\quad
\mbox{ for any }\vv,\,\x,\,\vec u\in V.
\ee

The finite set of vectors $\R\subset V$ is a {\it root system}
if $\R$ is ${R}_\vv$-invariant for any $\vv \in \R$ and
the group $W(\R)$ generated by all reflections ${R}_\vv$ with
$\vv \in \R$ (Coxeter group) is finite.

Let ${\cal H}^\alpha$ ($\alpha=0,1$) be two copies of $V$ with
orthonormal bases $a^\alpha_i$ ($i=1,\,...\,,\,N$), respectively.
For every vector $\vv=\sum_{i=1}^N v_i{\vec e_i}\in V$ let
$v^\alpha\in {\cal H}^\alpha$
be the vectors $v^\alpha=\sum_{i=1}^N v_i a^\alpha_i$,
so the bilinear forms on
${\cal H}^0\oplus {\cal H}^1$
can be defined as
\be
\label{bill}
(x^\alpha,\,y^\beta)=(\vec x,\, \vec y),
\ee
where $\vec x,\,\vec y \in V$ and $x^\alpha,\,y^\alpha \in {\cal
H}^\alpha$ are their copies.
The reflections $R_\vv$ act on ${\cal H}^\alpha$ as follows
\be
R_\vv(h^\alpha)=h^\alpha -
2\frac{(h^\alpha,\,v^\alpha)}{(\vv,\,\vv)}v^\alpha,
\qquad \mbox{ for any } h^\alpha\in {\cal H}^\alpha.
\ee
So the $W(\R)$-action on the spaces ${\cal H}^\alpha$
is defined.

Let $\nu$ be a set of constants $\nu_\vv$ with $\vv\in\R$ such that
$\nu_\vv=\nu_{\vec w}$ if $R_\vv$ and $R_{\vec w}$ belong to
one conjugacy class of $W(\R)$.
Consider the associative algebra $H_{W({R})}(\nu)$
of polynomials in the $a^\alpha_i$ with coefficients
in the group algebra ${\open C}[W(\R)]$ subject to the relations
\bee\label{rel1}
R_\vv h^\alpha=R_\vv(h^\alpha)R_\vv , \quad \mbox{ for any }\vv\in R,
                  \quad \mbox{ and } h^\alpha \in {\cal H}^\alpha \nn
\label{rel}
  [ h_1^\alpha, h_2^\beta] = \varepsilon^{\alpha\beta}
       \left((\vec h_1,\, \vec h_2)+
       \sum_{\vv\in\R} \nu_\vv
\frac {(\vec h_1,\,\vv)(\vec h_2,\,\vv)}{(\vv,\,\vv)}R_\vv\right)
\mbox{ for any  $h_1^\alpha$, $h_2^\alpha\, \in {\cal H}^\alpha$}.
\eee
where $\varepsilon^{\alpha\beta}$
is the antisymmetric tensor, $\varepsilon^{01}=1$.

This algebra
has faithful representation
via Dunkl differential-difference operators \cite{Dunkl} acting on
the space of infinitely smooth functions on $V$. Namely, let
\be\label{Dun}
D_i=
\di +\half \sum_{\vv\in\R} \nu_\vv\frac {v_i} {(\x,\,\vv)} (1-R_\vv)
\ee
and
\cite{Poly, BHV}
\be\label{aa}
a_i^\alpha =\frac 1 {\sqrt{2}} (x_i + (-1)^\alpha D_i),\quad
\alpha =0,1.
\ee

The reflections $R_\vv$ transform the deformed creation
and annihilation operators
(\ref{aa})
as vectors:
\be\label{comav}
R_\vv a_i^\alpha = \sum_{j=1}^N \left(\delta_{ij} - 2
\frac {v_i v_j}{(\vv,\,\vv)}\right)a_j^\alpha  R_\vv.
\ee
Since $[D_i,\, D_j]=0$ \cite{Dunkl}, it follows that
\be\label{comaa}
[a^\alpha_i, a^\beta_j] = \varepsilon^{\alpha\beta}
\left(\delta_{ij}+
\sum_{\vv\in\R} \nu_\vv \frac {v_i v_j}{(\vv,\,\vv)}R_\vv\right),
\ee
which manifestly coincides with (\ref{rel}).

We say that $H_{W({R})}(\nu)$ is {\it the algebra of
observables of Calogero model based on the root system $\R$}.

The commutation relations (\ref{rel}) suggest
to define the {\it parity} $\pi$ by setting:
\be\label{pi}
\pi (a_i^\alpha)=1\ \mbox{ for any }\alpha,\,i,
\qquad \pi(g)=0 \ \mbox{ for any } g\in W(\R)
\ee
and consider $H_{W({R})}(\nu)$ as a superalgebra.

Obviously, ${\open C}[W(\R)]$
is a subalgebra of $H_{W({R})}(\nu)$.

Observe an important property of superalgebra
$H_{W({R})}(\nu)$: the Lie superalgebra of its inner
derivations \footnote{Let ${\cal A}$ be arbitrary associative
superalgebra. Then, the operators ${\cal D}_x$
which act on ${\cal A}$ via
${\cal D}_x(y)=[x,\,y\}$ (supercommutator) constitute
the Lie superalgebra of {\it inner}
derivations.} contains  ${\fr sl}_2$ generated by
\be\label{sl2}
T^{\alpha\beta}=\half \sum_{i=1}^N
\left\{a_i^\alpha,\,a_i^\beta\right\}
\ee
which commute with ${\open C}[W(\R)]$, i.e.,
$[T^{\alpha\beta},\,R_\vv]=0$,
and act on $a_i^\alpha$ as on ${\fr sl}_2$-vectors:
\be\label{sl2vec}
\left[T^{\alpha\beta},\,a_i^\gamma\right]=
\varepsilon^{\alpha\gamma}a_i^\beta +
\varepsilon^{\beta\gamma}a_i^\alpha.
\ee

The restriction of operator
$T^{01}$ in the representation (\ref{aa}) on the subspace
of $W(\R)$-invariant functions on $V$
is a second-order differential operator which is the well-known
Hamiltonian of the rational Calogero model \cite{Cal} based on the
root system $\R$ \cite{OP}. The parameters $\nu_\vv$ are
the coupling constants of this model.
One of the relations (\ref{sl2}), namely,
$[T^{01},\,a_i^\alpha]=
-(-1)^\alpha a_i^\alpha$, allows one to find the wave functions
of
the equation $T^{01}\psi =\epsilon \psi$
via usual Fock procedure with
the vacuum $ |0\rangle$ such that
$a_i^0|0\rangle$=0 $\mbox{ for any } i$ \cite{BHV}.
After $W(\R)$-symmetrization these wave functions
become the wave functions of Calogero Hamiltonian.


\section{Supertraces on $H_{W({R})}(\nu)$.}\label{trace}

Any linear complex-valued function $str(\cdot)$ on the superalgebra
${\cal A}$ such that
\bee\label{scom}
str(fg)&=&(-1)^{\pi(f)\pi(g)}str(gf)
\eee
for any $f,g \in
{\cal A}$ with definite parity $\pi(f)$ and $\pi(g)$ is called a {\it
supertrace}.

Every supertrace $str(\cdot)$ on ${\cal A}$ generates
the invariant bilinear form on ${\cal A}$
\bee\label{bf}
B_{str}(f,g)=str(f\cdot g).
\eee

It is obvious that if such a bilinear form is degenerate,
then the
null-vectors (i.e., $v \in {\cal A}$ such that $B(v, x)=0$ for any
$x\in {\cal A}$) of
this form constitute the two-sided
ideal ${\cal I}\subset {\cal A}$.

The ideals of this sort are present
in the superalgebras $H_{W(A_1)}(\nu)$
(corresponding to the two-particle Calogero model) at $\nu =k+\half$
\cite{V} and in the superalgebras $H_{W(A_2)}(\nu)$
(corresponding to three-particle Calogero model)
at $\nu =k+\half$
and $\nu=k\pm\frac1 3$ \cite{K2}
for every integer $k$. For all the
other values of $\nu$ all supertraces on these superalgebras
generate the nondegenerate bilinear forms (\ref{bf}).

It is easy to describe all supertraces on
${\open C}[W(\R)]$. Every supertrace on ${\open C}[W(\R)]$
is completely determined by its values on
$W(\R)\subset {\open C}[W(\R)]$
and the function $str$ is a central function on $W(\R)$, i.e.,
the function constant on the conjugacy classes.

Before formulating the theorem establishing the connection between
the supertraces on $H_{W(\R)}(\nu)$ and the supertraces on
${\open C}[W(\R)]$,
let
us introduce the grading $E$ on the vector space of
${\open C}[W(\R)]$. Consider the subspaces
\be\label{Halpha}\label{eigs}
{\cal E}^\alpha (g) =\{h\in {\cal H}^\alpha:\quad gh=-hg \}
\mbox{ for } g\in W(\R).
\ee
Clearly, $\dim~{\cal E}^0(g)=\dim~{\cal E}^1(g)$.
Set \footnote{
It follows from Lemma 3 formulated below that $\rho(g)=E(g)|_{mod 2}$
is a grading on the group algebra ${\open C}[W(\R)]$.
It is well known parity of elements of the Coxeter group.}
\be\label{grad}
E(g)=\dim~\,{\cal E}^\alpha(g).
\ee
Obviously, $E(g)$ is equal to the number of $(-1)$ in the spectrum
of matrix $g$.\footnote{Indeed, let $\vec x \in V$ be an eigenvector
of orthogonal matrix $g\in W(\R)$, i.e., $g\vec x =\lambda\vec x $.
Then (\ref{rel1}) implies the relation
$gx^\alpha=\lambda^{-1}x^\alpha g$
in $H_{W(\R)}(\nu)$.}

The following theorem was proved in \cite{KV}
\footnote{This theorem was proved for the case $\R=A_N$ only
but the proof does not depend on the particular properties
of the symmetric group $S_N=W(A_{N-1})$.}:

{\bf Theorem 1.} {\it Let ${\cal P}(g)$ be the projection
${\open C}[W(\R)] \rightarrow {\open C}[W(\R)]$
defined as
\be \label{(*)}
{\cal P}(\sum_i \alpha_i
g_i)=\sum_{i:\, g_i \neq {\bf 1}} \alpha_i g_i
\mbox{ for $g_i\in W(\R)$, $\alpha_i\in {\open C}$}.
\ee
 Let the grading $E$ defined in (\ref{grad})
and the subspaces ${\cal E}^\alpha(g)$ defined in (\ref{Halpha})
satisfy the equations
\bee\label{main0}
E({\cal P}([h_0,\,h_1])g)=E(g)-1 \ \ \mbox{ for any } g\in
W(\R), \ \
\mbox{ and } h_\alpha\in {\cal E}^\alpha(g).
\eee
Then every supertrace on the algebra ${\open C}[W(\R)]$
satisfying the equations
\bee\label{GLC}
str([h_0,\,h_1]g)=0\qquad \mbox{ for any } g\in W(\R) \mbox{
with }E(g)\neq 0 \mbox{ and } h_\alpha \in {\cal E}^\alpha(g),
\eee
can be uniquely extended to a supertrace on $H_{W(\R)}(\nu)$.}

It is shown below that conditions (\ref{main0}) hold
for arbitrary Coxeter group $W(\R)$ and the number of
independent solutions of conditions (\ref{GLC}) is equal to
the number of conjugacy classes in $W(\R)$ with $E(g)=0$.


\section{Conditions \protect
(\ref{main0}) for an arbitrary Coxeter group.}

{\it Lemma 2. Let $g$ be an orthogonal $N\times N$ matrix
which has no eigenvalue $-1$,
i.e., the matrix $g+1$ is invertible.
Then the matrix $R_{\vv}g$ has exactly one
eigenvalue equal to $-1$.}

To prove this lemma let us consider the equation
$R_{\vv}g\x+\x=0$ or $g\x+ R_{\vv}\x=0$
for eigenvector $\x$ corresponding to eigenvalue $-1$.
Using the definition of $R_{\vv}$
one has $g\x+\x-2\frac {(\vv,\, \x)} {|\vv|^2} v =0$; hence,
$\x=2\frac {(\vv,\, \x)} {|\vv|^2}(g+1)^{-1} \vv$.
It remains to show that this equation has a nonzero solution.
Let $\vv=(g+1)\vec w$.
Then $|\vv|^2=2(|\vec w|^2+(\vec w,\, g\vec w))$ and
$((g+1)^{-1} \vv, \, \vv)= |\vec w|^2+(\vec w,\, g\vec w)$.
So the vector $\x_1=2\frac 1 {|\vv|^2}(g+1)^{-1} \vv$ is the only
(up to a factor) solution.

{\it Lemma 3. \label{l2} Let $g$ be an orthogonal $N\times N$ matrix
and $\cc_i$
($i=1,\, ... ,\, E(g)$) be the complete orthonormal set of its
eigenvectors corresponding to eigenvalue $-1$.
Then
\\{\em i)}
$E(R_{\vv} g) = E(g)+1$ if $(\vv,\,\cc_i)=0$ for all $i$;
\\{\em ii)}
if there exists an $i$ such that $(\vv,\, \cc_i) \neq 0$, then
$E(R_{\vv} g)=E(g)-1$ and the space of the
eigenvectors of $R_{\vv} g$ corresponding to
eigenvalue $-1$ is the subspace of
$span\{\cc_1,\,...,\, \cc_{E(g)}\}$ orthogonal to $\vv$.}

 Let $\cc_i$, $i=1,\,...\, N$, be the complete orthonormal set of
the eigenvectors of $g$, i.e. $g\cc_i=\lambda_i \cc_i$.
Here $\lambda_i=-1$ for $1\leq i\leq E(g)$. Let $x^i=(\x,\, \cc_i)$
for every vector $\x$. Consider the equation for the eigenvector
$\x=\sum_1^N x^i \cc_i$ corresponding to eigenvalue $-1$ of matrix
$R_{\vv} g$:
\be
\label{-1} (\lambda_i+1) x^i -2\frac {(g\x,\,\vv)} {|\vv|^2}v^i=0,
\ee
where $v^i=(\vv, \cc_i)$. It follows from (\ref{-1})
that either $v^i=0$
for $1\leq i \leq E(g)$ or $(g\x,\, \vv)=0$.
In the first case, one can consider the restriction of $R_{\vv}$
and $g$ onto the subspace spanned by $\cc_i$ with $i>E(g)$
and apply Lemma 2 to this
restriction and obtain {\em i)}.
In the second case, it follows from equation (\ref{-1}) that
$x^i=0$ for $i>E(g)$, hence, $g\x=-x$, $(\x, \,\vv)=0$
which yields {\em i)}.

Now one can prove the following

{\bf Theorem 4.} {\it Let $g\in W(\R)$.
Let
$c^\alpha_1, c^\alpha_2\in {\cal E}^\alpha (g)\subset H_{W(\R)}$
(i.e. $gc^\alpha_1=-c^\alpha_1g$, $gc^\alpha_2=-c^\alpha_2g$).
Let ${\cal P}(g)$ be the projection
(\ref{(*)}).
Then
\bee\label{main}
E({\cal P}([c^\alpha_1,\,c^\beta_2])g)=E(g)-1 \ \
\mbox{ for any } g\in W(\R).
\eee
}

Proof easily follows from the formula
\be
{\cal P}([c^\alpha_1,\,c^\beta_2])=\varepsilon^{\alpha\beta}
\sum_{\vv\in\R} \nu_\vv \frac {(\vec c_1,\,\vv)
(\vec c_2,\,\vv)}{(\vv,\,\vv)}R_\vv\,.
\ee
Indeed, if $(\vec c_1,\,\vv)(\vec c_2,\,\vv)\neq 0$, then
Lemma 3 implies that $E(R_\vv g)=E(g)-1$.


\section{The supertraces on ${\G}$, Ground Level Conditions
and the number of supertraces on $H_{W(\R)}(\nu)$.}
\label{Q_N}

Due to the $W(\R)$-invariance, the definition of the supertrace
on $\G$ is the definition of the central function on $\G$ i.e.
a function constant on each conjugacy class of $\G$.
Thus the number of the supertraces on $\G$
is equal to the number of conjugacy classes in $\G$.

Since $\G \subset H_{W(\R)}(\nu)$, some
additional restrictions on these functions
follow from (\ref{scom}) and the defining relations (\ref{rel}) for
$H_{W(\R)}(\nu )$.  Indeed, consider some elements $c_i$ such that
$gc_i=-c_ig$,
where $g\in {W(\R)}$ and $c_i\in {\cal H}^0 \oplus {\cal H}^1$.
Then, one finds from (\ref{scom}) and (\ref{eigs}) that
$str \left ( c_i c_j g
\right )$= $ - str \left ( c_j g c_i\right )$= $ str \left ( c_j c_i
g \right )$
and, therefore, $ str \left ( [ c_i, c_j] g \right )=0 $.

Since $[ c_i, c_j] g \in \G $,
these conditions restrict supertraces of degree-0 polynomials in
$a^\alpha_i$.
In \cite{KV} we called them Ground Level Conditions (GLC).

They express the supertrace of elements $g$ with $E(g)=e$
via the supertraces of elements $R_{\vv}g$ with $E(R_{\vv}g)=e-1$:
\bee\label{GLC2}
str(g)=-str(([c^0_i,\, c^1_i]-1)g),
\mbox{ if } (\vec c_i,\,\vec c_i)=1.
\eee

Ground Level Conditions (\ref{GLC}) is an overdetermined system
of linear equations for the central functions on $\G$.

Let us prove by induction on $E(g)$ the following theorem

{\bf Theorem 5.}
{\it GLC (\ref{GLC}) have nonzero solutions
and the number of independent solutions is equal to the
number of conjugacy classes in $W(\R)$ with $E(g)=0$.}

The first step is simple: if $E(g)=0$, then $str(g)$ is an arbitrary
central function.
The next step is also simple: if $E(G)=1$, then there exists
a unique element $c^0_1\in {\cal E}^0(g)$
and a unique element $c^1_1\in {\cal E}^1(g)$ such that
$|c^\alpha_1|=1$ and $gc^\alpha_1=-c^\alpha_1g$.
Since $(([c^0_1,\, c^1_1]-1)g)\in \G$ and
$E(([c^0_1,\, c^1_1]-1)g)=0$, then
\be\label{sol}
str(g)=-str(([c^0_1,\, c^1_1]-1)g)
\ee
is the unique possible value for $str(g)$ with $E(g)=1$.
A priori these values are not consistent with other GLC.

Suppose that Ground Level Conditions
\be
str \left ( [ c^0_i, c^1_j] g \right ) =0
\ee
considered for all $g$ with $E(g)\leq e$
and for all $c^\alpha_i\in {\cal E}^\alpha (g)$
($i=1,,\,...\,,e$) such that $(c^\alpha_i,\,c^\beta_j)=\delta_{ij}$
have $Q_e$ independent solutions.

{\bf Statement 6} {\it The value $Q_e$ does not depend on $e$.}

It was shown above that $Q_1=Q_0$. Let $e\geq 1$.
Let us consider $g\in W(\R)$ with $E(g)=e+1$.
Let $c^\alpha_i\in {\cal E}^\alpha(g)$ ($i=1,2$) be such that
$(c^\alpha_i,\,c^\beta_j)=\delta_{ij}$.
These elements give the conditions:
\bee
\label{e11} str(g)=-str(([c^0_1,\, c^1_1]-1)g), \\
\label{e22} str(g)=-str(([c^0_2,\, c^1_2]-1)g), \\
\label{e21} str([c^0_1,\, c^1_2]g)=0.
\eee
Let us transform (\ref{e11}):
\bee
str(g)&=&str(S_1)-str(S_{12}) \mbox{, where}\\
S_1&=&-\left([c^0_1,\, c^1_1]-1
  -\sum_{\vv\in\R:\,(\vv,\,\vec c_1)(\vv,\,\vec c_2)\neq 0}
\nu_\vv \frac {(\vv,\,\vec c_1)^2}{|\vv|^2}
   R_\vv \right) g =
\nn
&{}&
  -\left(\sum_{\vv\in\R:\,\,(\vv,\,\vec c_1)(\vv,\,\vec c_2)=0}
\nu_\vv \frac {(\vv,\,\vec c_1)^2}{|\vv|^2}
   R_\vv \right) g =
\nn
\label{s1}
&{}&
  -\left(\sum_{\vv\in\R:\,\,(\vv,\,\vec c_2)=0}\nu_\vv
\frac {(\vv,\,\vec c_1)^2}{|\vv|^2}
   R_\vv \right) g, \\
S_{12}&=& \left(
   \sum_{\vv\in\R:\,(\vv,\,\vec c_1)(\vv,\,\vec c_2)\neq 0}
\nu_\vv \frac {(\vv,\,\vec c_1)^2}{|\vv|^2}R_\vv
   \right)g.
\eee
It is evident from (\ref{s1}) and Lemma 3 that $E(S_1)=e$ and
$S_1c^0_2=-c^0_2S_1$.
Hence, due to (\ref{GLC2}) and inductive hypothesis
\be
str(S_1)=-str(([c^0_2,\,c_2^1]-1)S_1)=
str(([c^0_2,\,c_2^1]-1) (([c^0_1,\,c_1^1]-1)g - S_{12}))
\ee
and as a result
\be
str(S_1)=
str(([c^0_2,\,c_2^1]-1)([c^0_1,\,c_1^1]-1)g )
-str(([c^0_2,\,c_2^1]) S_{12})
+str(S_{12}).
\ee
Finally, (\ref{e11}) is equivalent under inductive hypothesis to
\be\label{e11a}
str(g)=
str(([c^0_2,\,c_2^1]-1)([c^0_1,\,c_1^1]-1)g )
-str(([c^0_2,\,c_2^1]) S_{12}).
\ee
Analogously, (\ref{e22}) is equivalent under inductive hypothesis to
\be\label{e22a}
str(g)=
str(([c^0_1,\,c_1^1]-1)([c^0_2,\,c_2^1]-1)g )
-str(([c^0_1,\,c_1^1]) S_{21}),
\ee
where
\be
S_{21} = \left(
   \sum_{\vv\in\R:\,(\vv,\,\vec c_1)(\vv,\,\vec c_2)\neq 0}
\nu_\vv \frac {(\vv,\,\vec c_2)^2}{|\vv|^2}R_\vv
   \right)g.
\ee
Now, let us compare the corresponding terms in (\ref{e11a}) and
(\ref{e22a}).
First, the relation
\be
str(([c^0_1,\,c_1^1]-1)([c^0_2,\,c_2^1]-1)g )=
str(([c^0_2,\,c_2^1]-1)([c^0_1,\,c_1^1]-1)g )
\ee
is identically true for every (super)trace on ${\open C}[W(\R)]$, as
$[c^0_1,\,c_1^1]$ commutes with $g$.
Second,
\be
str(([c^0_1,\,c_1^1]) S_{21})=
str(([c^0_2,\,c_2^1]) S_{12})
\ee
since
\be\label{r12}
str( [c^0_1,\,c_1^1] (\vv,\,\vec c_2)^2 R_\vv g)=
str( [c^0_2,\,c_2^1] (\vv,\,\vec c_1)^2 R_\vv g)
\ee
for every $\vv\in\R$ such that
$(\vv,\,\vec c_1)(\vv,\,\vec c_2)\neq 0$.
Indeed, due to Lemma 3 the element
\be
\vec c={\beta_1} \vec c_1 + {\beta_2} \vec c_2\,,\mbox{ where }
{\beta_1}=-(\vv,\,\vec c_2)\neq 0 \mbox{ and }
{\beta_2}=(\vv,\,\vec c_1)\neq 0\,,
\ee
is orthogonal to $\vv$:
\be\label{ort1}
(\vv,\,\vec c)=0
\ee
and satisfies the relation
\be
R_\vv g c^\alpha = - c^\alpha R_\vv g
\ee
due to Lemma 3. This fact together with
\be
E({\cal P}([c^0_i,\,c^1]) R_\vv g) = e-1 \mbox{ for } i=1,2
\ee
(this also follows from Lemma 3)
and inductive hypothesis imply
\be\label{ort2}
str([c^0_i, \,c^1] R_\vv g)=
str([c^0, \,c^1_i] R_\vv g)=0 \quad (i=1,2).
\ee
Substituting
$\vec c_1=\frac 1 {\beta_1} (\vec c-{\beta_2} \vec c_2)$
and $\vec c_2=\frac 1 {\beta_2} (\vec c-{\beta_1} \vec c_1)$
in the left-hand side of (\ref{r12})
and using (\ref{ort1}) and (\ref{ort2})
one obtains the right-hand side of (\ref{r12}).
Thus, (\ref{e11}) is equivalent to (\ref{e22});
hence
\be
str(([c^0_1,\, c^1_1]-1)g)-str(([c^0_2,\, c^1_2]-1)g)=0
\ee
for every orthonormal pair $c_1,\,c_2\in {\cal E} (g)$.
Consequently,
\be
str([c^0_1,\, c^1_2]g)=0
\ee
which finishes the proof of Statement 6 and Theorem 5.



\begin{thebibliography}{99}
\bibitem{Dunkl} C.F.Dunkl, Trans. Am. Math. Soc. {\bf 311} (1989) 167.
\bibitem{Poly} A.~Polychronakos, Phys.~Rev.~Lett. {\bf 69} (1992) 703.
\bibitem{BHV} L.~Brink, H.~Hansson and M.A.~Vasiliev,
             Phys.~Lett. {\bf B286} (1992) 109.
\bibitem{Cal} F.~Calogero, J.~Math.~Phys., {\bf 10} (1969) 2191, 2197;
            {\it ibid} {\bf 12} (1971) 419.
\bibitem{OP} M. A.Olshanetsky and A.~M. Perelomov,
              Phys. Rep., {\bf 94} (1983) 313.
\bibitem{V} M.A.~Vasiliev, JETP Letters, {\bf 50} (1989) 344-347;
              Int. J. Mod. Phys. {\bf A6} (1991) 1115.
\bibitem{K2} S.E.Konstein,
              Teor.~Mat.~Fiz., {\bf 116} (1998) 122, hep-th/9803213.
\bibitem{KV} S.E.~Konstein and M.A.~Vasiliev,
              J.~Math.~Phys. {\bf 37} (1996) 2872, hep-th/9512038.
\end{thebibliography}
\end{document}